\documentclass[preprint,showpacs,preprintnumbers,amsmath,amssymb,twoside,aps]{revtex4}

\usepackage{graphicx}
\usepackage{dcolumn}
\usepackage{bm}
\usepackage{color}

\begin{document}
\newcommand{\NCOx}{Na$_{\mbox{\scriptsize x}}$CoO$_{\mbox{\scriptsize 2}}$}
\newcommand{\NCO}{Na$_{\mbox{\scriptsize 0.73}}$CoO$_{\mbox{\scriptsize 2}}$}
\newcommand{\NCOc}{Na$_{\mbox{\scriptsize 0.75}}$CoO$_{\mbox{\scriptsize 2}}$}
\newcommand{\tco}{T$_{\mbox{\scriptsize CO}}$}
\newcommand{\tso}{T$_{\mbox{\scriptsize SO}}$}
\newcommand{\tc}{T$_{\mbox{\scriptsize C}}$}
\newcommand{\tilt}{I$_{\mbox{\scriptsize 300}}$}
\newcommand{\GM}{$\Gamma$M}
\newcommand{\GK}{$\Gamma$K}
\newcommand{\LPSMO}{(La$_{\mbox{\scriptsize 1-y}}$Pr$_{\mbox{\scriptsize y}}$)$_{\mbox{\scriptsize 7/8}}$Sr$_{\mbox{\scriptsize 1/8}}$MnO$_{\mbox{\scriptsize 3}}$}
\newcommand{\quadru}{ $(1\pm 0.25,\pm 0.25,0)$}
\newcommand{\doub}{ $(1\pm 0.5,\pm 0.5,0)$}

%


\title{Anisotropic quasiparticle renormalization in Na$_{0.73}$CoO$_2$:
 role of inter-orbital interactions and magnetic correlations}

\author{J. Geck$^{1,2,}$\footnote{Email: geck@physics.ubc.ca}}
\author{S.V. Borisenko$^1$}
\author{H. Berger$^3$}
\author{H. Eschrig$^1$}
\author{J. Fink$^1$}
\author{M. Knupfer$^1$}
\author{K. Koepernik$^1$}
\author{A. Koitzsch$^1$}
\author{A.A. Kordyuk$^{1,4}$}
\author{V.B. Zabolotnyy$^1$}
\author{B. B\"uchner$^1$}

%
%

\affiliation{$^1$IFW Dresden, P.O. Box 270116,D-01171 Dresden, Germany}

\affiliation{$^2$Department of Physics and Astronomy, University of British Columbia, Vancouver, BC, V6T 1Z1, Canada}

\affiliation{$^3$Institut de physique de la mati\'{e}re complex, EPF Lausannne, 1015 Lausanne, Switzerland}


\affiliation{$^4$Institute of Metal Physics of the National Academy of Sciences of Ukraine, 03142 Kyiv, Ukraine}

\date{Received: \today}
%
\begin{abstract}
We report an angular resolved photoemission study of \NCOx \/ with x$\simeq$0.73 where it is found that the renormalization of the quasiparticle
dispersion changes dramatically upon a rotation from \GM \/ to \GK. The comparison of the experimental data to the calculated band structure reveals
that the  quasiparticle renormalization is most pronounced along the \GK-direction, while it is significantly weaker along the \GM-direction. We
discuss the observed anisotropy in terms of multiorbital effects and point out the relevance of magnetic correlations for the band structure of \NCOx
\/ with $x\simeq0.75$.
\end{abstract}

\maketitle


The unconventional behavior of correlated electrons in materials comprised of square lattices has attracted a vast amount of attention
\cite{OrensteinScience00,TokuraScience00}. However, besides strong electronic correlations, the topology of the underlying lattice structure itself
constitutes another important ingredient that can produce exotic electronic ground states.

The CoO$_2$-layers in the \NCOx\/ materials, which are built up from edge-sharing CoO$_6$ octahedra, constitute a realization of a correlated
electron system based on a triangular lattice. More specifically, these compounds possess a layered hexagonal structure, where strongly covalent
CoO$_2$- and ionic Na-layers ($ab$-planes) alternate along the perpendicular $c$-axis \cite{HuangPRB04a}. Upon changing the Na content $x$, these
materials can be doped with electrons and, in addition, water molecules can be intercalated. Both changing $x$ and water intercalation drastically
alter the electronic properties of \NCOx, leading most notably to the emergence of superconductivity upon hydration \cite{FooPRL04,TakadaNature03}.
We will focus on the non-hydrated compounds with $x\simeq0.7$, where an anomalous metallic state with an extremely large and field dependent
thermoelectric power as well as a giant field dependent scattering rate was observed \cite{WangNature03,LiPRL04}. There is also evidence for the
seemingly paradoxical coexistence of electron itinerancy and localized magnetic moments, as well as for unusual charge order CDW phenomena in these
macroscopically metallic compounds \cite{GavilanoPRB04,BernhardPRL04,NingPRL04}.

%
The unconventional electronic properties described above strongly motivate the study of the electronic structure of \NCOx\/ by means of angular
resolved photoemission spectroscopy (ARPES), which provides direct and unique experimental access to single-particle excitations and, thus, to the
many-body effects in these materials. Previous ARPES studies on \NCOx\/ \cite{VallaNature02,YangPRL04,HasanPRL04,YangPRL05,QianPRL06,QianPRL06b}
showed pronounced deviations from the electronic structure predicted by LDA
calculations \cite{SinghPRB00}. 
In addition, a strongly renormalized heavy QP band, displaying a  kink feature was reported \cite{YangPRL04,HasanPRL04,YangPRL05}.

In this ARPES study we focus on the momentum ($k$) dependent renormalization effects in \NCO , which is
essential to unravel the relevant couplings that govern the low energy physics. 
%
ARPES experiments were carried out using a lab-based system equipped with a SCIENTA SES 200 analyser and a Gammadata He discharge lamp at an
excitation energy $h \nu = 21.2$\,eV (energy resolution 30\,meV, angular resolution 0.3$^{\circ}$).
The high-quality \NCO\/ single crystals examined in this study were grown by the sodium chloride flux methods as described in Ref.\,\cite{Iljev04}.



A typical Fermi level ($E_F$) crossing observed at T=25\,K is shown in Fig.\,1, where the image on the left hand side shows the photoelectron
intensity as a function of momentum $k$ and binding energy $E_B$. A single and well defined QP band, which crosses the Fermi energy at $E_B=0$\,eV
can be observed. On the right hand side of Fig.\,1 an energy distribution curve (EDC) and a momentum distribution curve (MDC)
are shown. 
%
By mapping the $E_F$ crossings over a large area of $k$-space, a cut through the Fermi surface (FS) of \NCO\/ parallel to the CoO$_2$ planes was
measured. The data allows for a precise calibration of the $k$-scale, because dispersions from the first into the second Brillouin zone were
captured. In Figs.\,2\,(a),(b) the $k$-dependent photoelectron intensity for two fixed binding energies of $E_B=0$\,eV and 0.132\,eV are shown.
Focussing on the cut at $E_B=0$\,eV in Fig.\,2\,(a), a large hole-pocket centered around the Brillouin zone center $\Gamma$ is observed. The FS
displays a clear hexagonal topology. The intensity variations along the FS, leading to a seemingly lower symmetry, are caused by matrix element
effects.
The size of the observed FS agrees well with the doping level, provided that only one of the two $a_{1g}$-bands (cf. Fig.\,2\,(c)) crosses $E_F$.
This agrees with the data shown in Fig.\,1, where a single QP-band crosses $E_F$ .

%

In accordance with previous studies, the hole-pockets along \GK\/ are not observed for $E_B=0$\,eV.  It can be seen in Fig.\,2\,(b) that the
corresponding $e_g'$-band (cf. Fig.\,2\,(c)) lies well below $E_F$, as reported earlier \cite{YangPRL05}.
This can also be observed in Fig.\,2\,(c), where the measured band structure along \GK \/ and \GM \/ is compared to the band structure obtained by
LDA \cite{SinghPRB00}. There is a fair agreement of the ARPES data and LDA at higher binding energies above 1.5\,eV. However, close to $E_F$
deviations occur which are particularly pronounced along \GK.

In Fig.\,3 two cuts of the data set shown in Fig.\,2 along the \GK- (left) and \GM-direction (right) are shown.
Focussing on the cut along  \GK , two prominent features can be observed. First, there is a band crossing $E_F$ that forms the FS. Second, there is
another band at higher binding energies, which can be identified as the $e_g'$-band. The data indicates that the top of the $e_g'$-band is at about
85\,meV.

By tracking the maximum of the MDCs for different $E_B$, the QP-dispersion ($E(k)$) indicated by black symbols is obtained. Close to $E_F$ the
dispersion can be well described by a linear behavior, yielding the Fermi velocity  $v_F=(0.3\pm0.05)$\,eV\,\AA \/ along \GK .
However, at slightly higher $E_B$, the measured dispersion is bent and thus deviates from the linear behavior close to $E_F$.
We define the energy where this deviation becomes significant as $E_d$, resulting in $E_d=(26\pm 8)$\,meV for the \GK-direction. The second bend in
the dispersion around 80\,meV is related to the crossing of the two bands \cite{QianPRL06b}.
Applying the same analysis to the cut along \GM\/ leads to $v_F=(0.6\pm0.08)$\,eV\,\AA\/ and $E_d=(66 \pm 5)$\,meV. Clearly, both $v_F$ and $E_d$
depend on the direction in $k$-space.

The observed bending of the dispersion that sets in around $E_d$ could be due to a coupling of the QP to bosonic excitations. In fact, consistent
with this interpretation, we observe an enhanced increase of the scattering rate around $E_d$.
It has to be noted, however, that the feature in the dispersion observed here is not as well defined as the kink in the cuprates, for example. This
means that  the values of $E_d$ cannot be identified straightforwardly with a specific mode energy and, moreover, a coupling to several different
bosonic excitations is possible.

In the following we will focus on $v_F$, which is a well defined quantity: Cuts for different $\psi$-values (cf. Fig.\,1\,(b)) were systematically
analyzed in the same way as described above. The obtained variation of $v_F$ as determined from the ARPES data is shown in Fig.\,4\,(a). Upon
rotating the direction of the cut from \GM\/ to \GK\/, $v_F$ decreases by about a factor of 2.
Using the function $k_F(\psi)$ that was determined from the data in Fig.\,2\,(a) together with the obtained values of $v_F$, the variation of the
effective mass $m^*$ with $\psi$ can be calculated. The result is given in Fig.\,4\,(c). Remarkably, the QP along the \GK\/ direction is about twice
as heavy as the QP along \GM. This pronounced anisotropy of $m^*$ is expected to have a strong impact on the in-plane electronic properties of \NCO.

To determine the renormalization effects in \NCO, we use the LDA band structure as a reference and compare the ARPES Fermi velocities ($v_F^{PES}$)
to the corresponding LDA values ($v_F^{LDA}$). As it can be observed in Figs.\,4\,(a),(b), these two quantities show exactly the opposite behavior:
$v_F^{PES}$ decreases and $v_F^{LDA}$ increases upon rotation from \GM\/ to \GK.
In order to check whether the deviation between ARPES and LDA is related to a lattice distortion at the surface, LDA calculations were performed for
structures where the distance between the oxygen and the cobalt layers, i.e. the Co--O--Co bond angle, was changed. In agreement with previous
calculations, we observe that the top of the $e_g'$ band is shifted to higher binding energies upon increasing the Co--O--Co bond angle. At the same
time the anisotropy of $v_F^{\rm LDA}$ is slightly reduced, but unchanged qualitatively. This strongly suggests  that the measured anisotropy of
$v_F^{PES}$ is not caused by a lattice distortion at the surface.

 The comparison of LDA and ARPES therefore shows that the deviation of the LDA
bandstructure from the measured QP-dispersion increases dramatically close to the \GK-direction (cf. inset of Fig.\,3). In the following we will
refer to this deviation as QP-renormalization (QPR).
This QPR can be characterized using a constant $\kappa$ defined by $(1+\kappa)\,v_F^{PES}=v_F^{LDA}$. The $\psi$- or, in other words, k-dependence of
$\kappa$ is shown in Fig.\,4\,(d), revealing the strong anisotropy of the QPR in \NCO. We note that, although the photoelectron intensity in
Fig.\,2\,(a) at $k_F$ is also influenced by matrix element effects, it is always significantly lower in the \GK- than in the \GM-direction, in
agreement with enhanced renormalization effects along \GK.

We find $\partial_k E^{PES}\simeq \partial_k E^{LDA}/2$ along \GK \/ in the whole energy range up to $E_B=85$\,meV (inset of Fig.\,3).
%
At the same time it is remarkable that the QPR gets stronger the closer the so-called $e_g'$-band gets to the Fermi level (c.f. Fig.2\,(b)). This
points to an effect related to coupling between the $a_{1g}$- and the $e_g'$-bands. In fact, a strong interaction between these bands is manifested
by a large hybridization gap at higher binding energies  and the polarization dependence along \GK\/ found in a recent ARPES study \cite{QianPRL06b}.
Hence, the $k$-dependent QPR at $E_F$ is most likely caused by multiorbital effects, i.e. interactions between the states of $e_g'$ and $a_{1g}$
symmetry. In this case, the QP-states along \GM\/ and \GK\/ display different properties: Along \GM\/ the QP-states have largely $a_{1g}$ symmetry,
while they display pronounced multiorbital properties along \GK . This is of crucial importance for the many-body effects in these materials, since
the coupling of the QP to bosonic excitations depends critically on the symmetry of the QP-states \cite{DevereauxPRL04}.
%
To conclude so far, the observed anisotropies clearly indicate that multiorbital effects play an important role for the QP-dynamics at $E_F$.


Furthermore, our DFT studies --details will be provided in a forthcoming publication-- show that 
magnetic correlations play an important role for the QP-dynamics as well: According to non-magnetic LDA calculations the band structure of \NCOc\/
displays a strong 3D character. In agreement with a previous DFT study \cite{JohannesEPL04}, we obtain a sizeable $k_z$-dispersion parallel to the
$c$-axis that leads to additional caps of the FS as shown in Figs.\,5\,(a),(b). Such a strong 3D character is not in agreement with ARPES data: (i)
in general, the QP peaks at $E_F$ are expected to be considerably broadened in a 3D system, in particular because the short life time of the final
states becomes important \cite{StarnbergPRB93}. This is not the case (Fig.\,1). (ii) ARPES measurements at various excitation energies do not show
any evidence for a strong dispersion along $c$ \cite{QianPRL06b}.

However, magnetic LSDA calculations yield an AFM ground state, where ferromagnetic $ab$-planes are coupled antiferromagnetically along $c$. This
agrees well with neutron data \cite{HelmePRL05}. In the AFM state, the $k_z$-dispersion is strongly reduced, which removes the aforementioned FS-caps
and yields the FS shown in Fig.\,5\,(c). In other words, according to LSDA, 3D AFM correlations
render the electronic structure of \NCOc\/ effectively 2D. The top of the $e_g'$-band at $E_B\simeq 70$\,meV as well as the topology and size of the
FS obtained in LSDA are in good agreement with the ARPES data as demonstrated in 5\,(d). The above results together with the neutron data indicate
that AFM correlations have a strong influence on the electronic structure of \NCOx\/ with $x\simeq0.75$.

In conclusion, we have shown that the QPR in \NCO \/ is strongly anisotropic and provided clear evidence for the relevance of multiorbital effects
for the QP dynamics in this material. In addition, detailed DFT studies highlight the impact of magnetic correlations on the QP-states near $E_F$,
which is expected to be directly related to the unusual temperature as well as the field dependencies of the thermopower and the QP scattering rates
\cite{WangNature03,LiPRL04,HasanPRL04}.  Hence, both the interactions between the $a_{1g}$ and $e_g'$ states as well as magnetic correlations have to
be taken into account in order to obtain a realistic description of these materials.

{\bf Acknowledgements:} We thank Dr. Bussy (Univ. of Lausanne) for the micro probe analysis and  I. Elfimov, K.M. Shen, D.G. Hawthorn and G.A.
Sawatzky for helpful discussions. This work was supported by the Swiss NCCR research pool MaNEP of the Swiss NSF, the DFG (FOR 538 research unit,
Grant 51195121) and the BMBF (Grant 05KS4OD2/8). J.G. gratefully acknowledges the support by the DFG.

\section*{\large Figure captions}

{\bf FIG.\,1:} Left: Typical Fermi level crossing observed along a cut close to \GM\/ ($\psi=173^{\circ}$, cf. Fig.\,2\,(b))
Right: Corresponding energy distribution curve (EDC) and momentum distribution curve (MDC) at $k=k_F$ and $E_B=0$\,eV, respectively.

{\bf FIG.\,2:} ARPES data for \NCO\/ (excitation energy $h \nu=21.2$\,eV). (a), (b): Momentum distribution maps of the photoelectron intensity
integrated over a small energy interval ($E_B\pm 3$\,meV) at $E_B=0$\,eV and 0.132\,eV measured at T=25\,K. The measured $k$-region is indicated by
the black dotted line in (a). The other regions in $k$-space have be obtained by rotating this data set by 120$^{\circ}$ and 240$^{\circ}$. The
broken white lines show the two-dimensional Brillouin zone. High-symmetry points $\Gamma$, K, and M are indicated in (a) and the definition of $\psi$
is given in (b). A fit to $k_F=k_F(\psi)$ is shown as a solid black line.  (c): Comparison of the measured band structure and the LDA calculation by
Singh (black lines) \cite{SinghPRB00}. The crystal field split $e_g'$- and $a_{1g}$-manifolds are indicated.\label{fig:2}

{\bf FIG.\,3:} Cuts through the map data shown in Fig.\,2\,(a) and (b). The data is normalized to binding energies above 0.25\,eV. Black symbols:
QP-dispersion determined by fitting MDCs at different $E_B$. Broken lines indicate the fitted linear dispersions (see text).  The inset shows the
ARPES- (symbols) and LDA- (lines) dispersions as a function of $k-k_F(\psi)$. LDA for $x=0.73$ in the rigid band approximation (cf. Fig.\,4).
\label{fig:3}

{\bf FIG.\,4:} (a),(b): $v_F^{PES}$ and $v_F^{LDA}$ as a function of $\psi$. The experimental $v_F^{PES}$ values at a given value of $\psi$ were
obtained by averaging over two equivalent cuts (e.g. $\psi=150^{\circ}$, $210^{\circ}$).  The LDA calculations were performed in the rigid band
approximation for the low temperature lattice structure using Wien2K. The same behavior was also found by LDA/LSDA calculations in the virtual
crystal approximation (cf. Fig.\,5). (c): Effective mass of the QP. (a)-(c): $h \nu=21.2$\,eV. Solid curves are fits to a sinus-function intended to
serve as guides to the eye. (d): $\kappa=v_F^{LDA}/v_F^{PES}-1$ characterizing the QPR. \label{fig:4}

{\bf FIG.\,5:} FS obtained for $x=0.75$ by LDA in the virtual crystal approximation (VCA), revealing a three-dimensional band structure. (c): FS for
the AFM ground state obtained by LSDA in the VCA where the band structure retains its pronounced two-dimensionality. The color scale in (a)-(c)
indicates $v_F$. (d) Comparison of the measured and LSDA FS. The DFT calculations have been performed using the FPLO code
\cite{EschrigPRB99}.\label{fig:5}
%


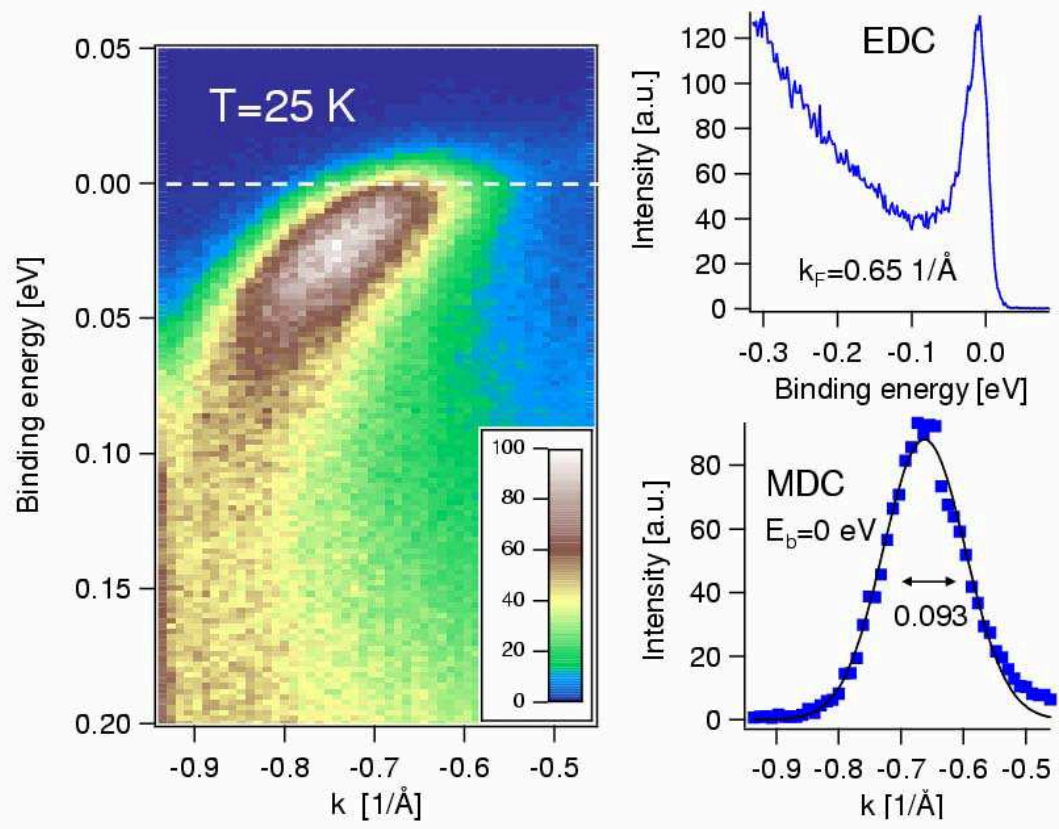
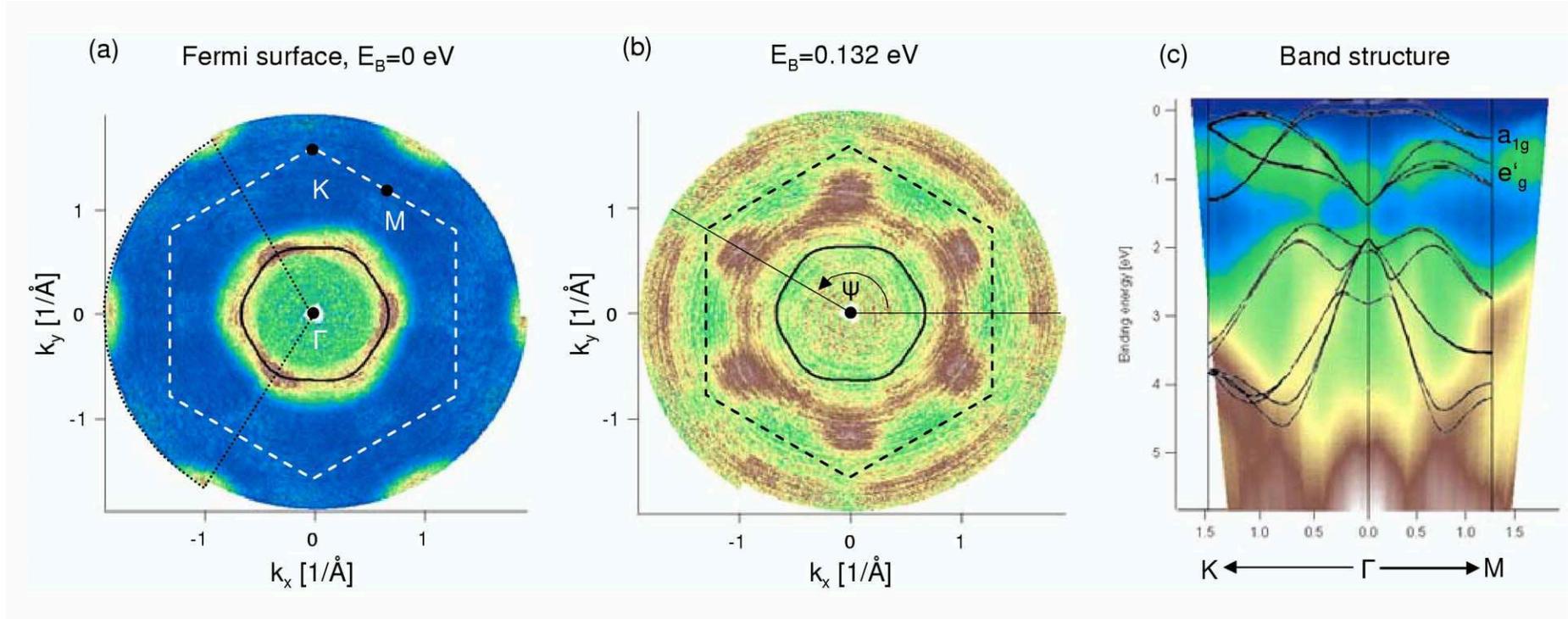
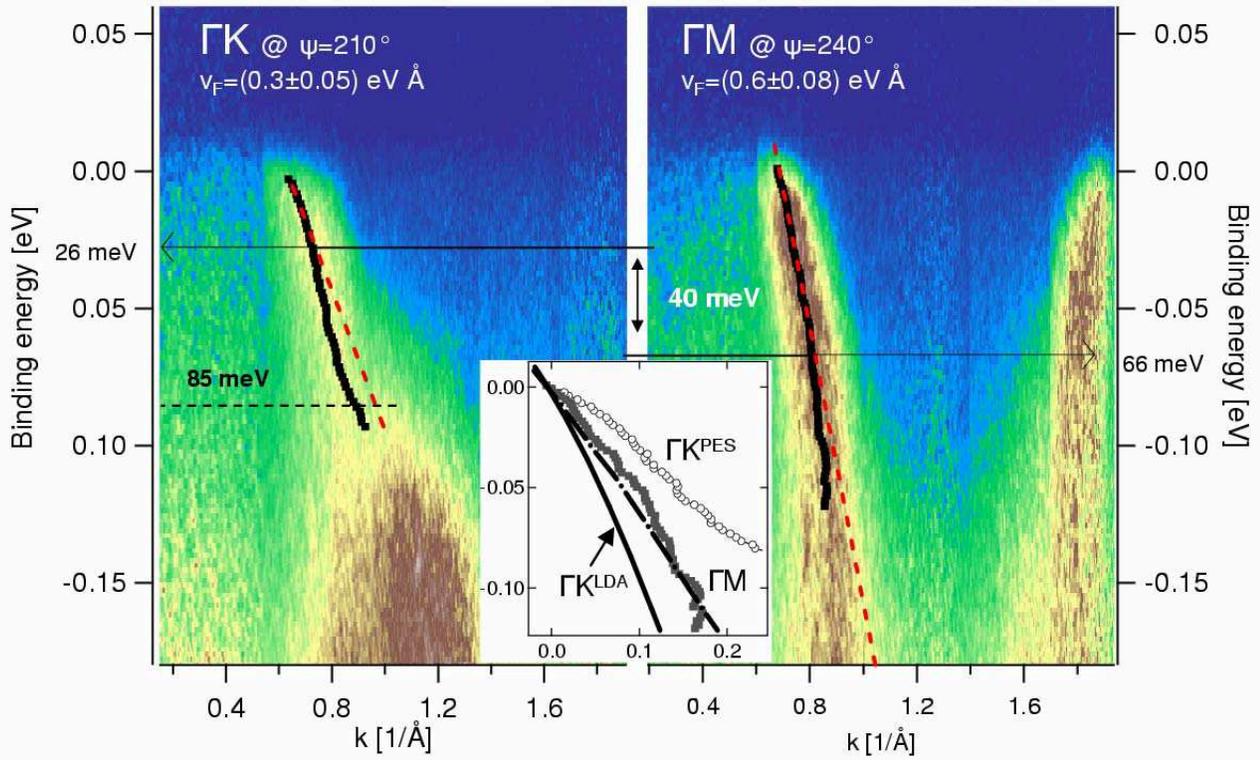
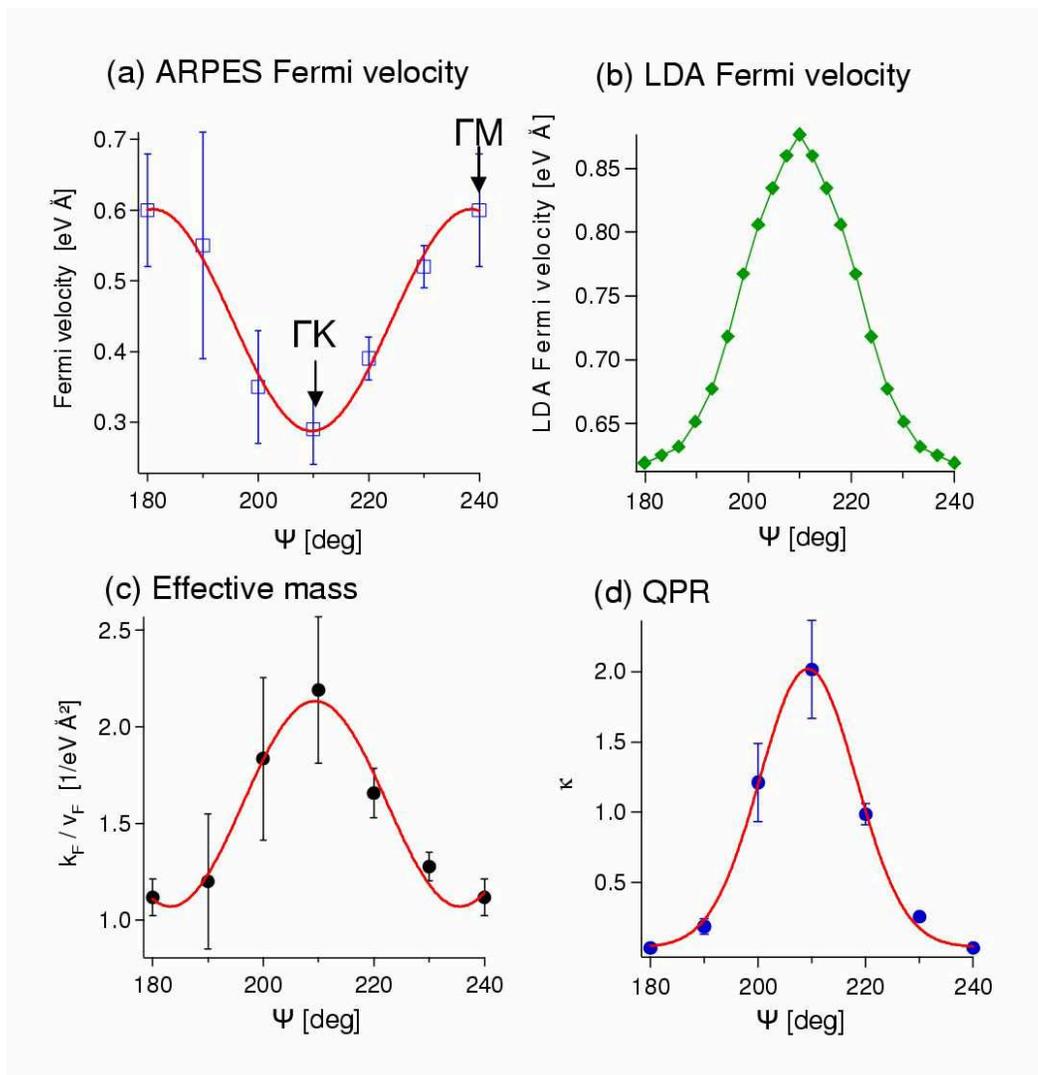
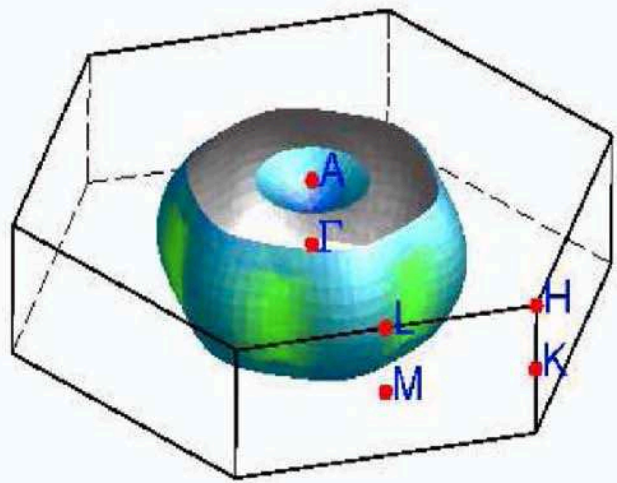
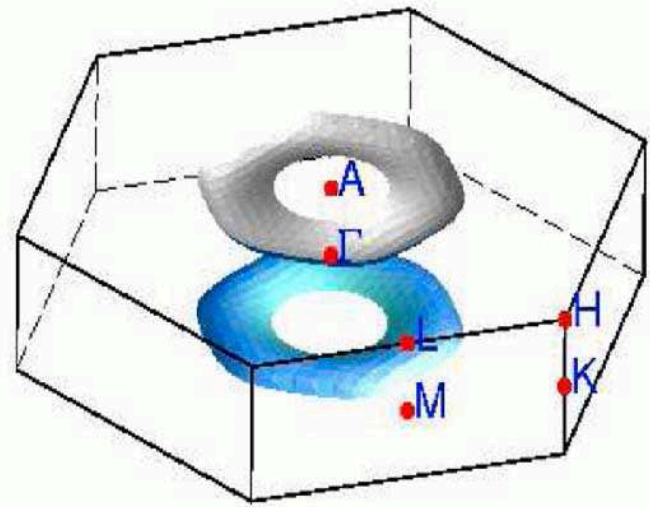
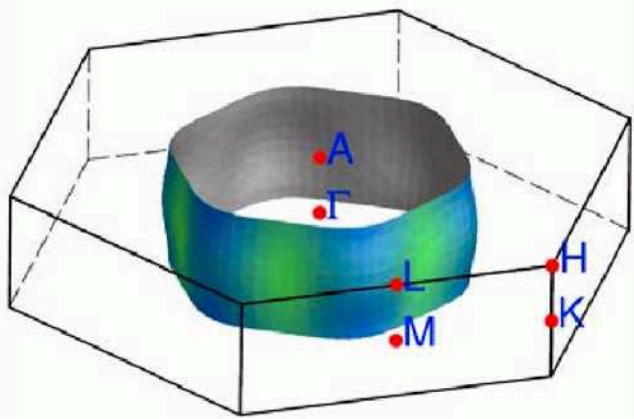
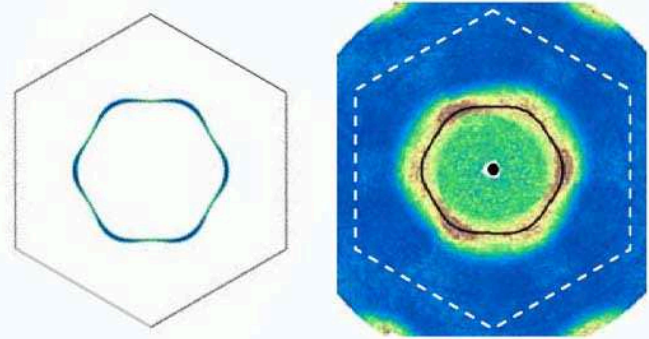
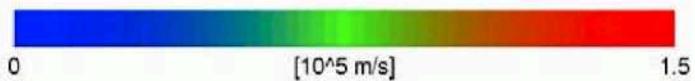
\end{document}